\documentclass[prl,nobibnotes,floatfix,superscriptaddress,twocolumn]{revtex4-1}
\usepackage{amsfonts,amsmath,graphicx,epsfig,latexsym}
\usepackage{subfigure}
\usepackage{graphicx}
\usepackage{color}
\usepackage{natbib}
\usepackage{multirow}
\usepackage{latexsym,amssymb}
\usepackage{amsmath}
\usepackage{soul}
\usepackage{ulem}
\usepackage[linktocpage,bookmarksopen,bookmarksnumbered]{hyperref}
\usepackage[dvipsnames]{xcolor}
\colorlet{LightRubineRed}{RubineRed!70!}
\colorlet{Mycolor1}{green!10!orange!90!}
\definecolor{Mycolor2}{HTML}{00F9DE}

\usepackage{comment}
\usepackage{braket}

\renewcommand{\vr}{{\mathbf{r}}} 
\newcommand{\vk}{{\mathbf{k}}}

 \newcommand{\vq}{{\mathbf{q}}}

\usepackage{multirow}
\usepackage[left]{lineno}

\begin{document}
\title{Electronic correlation in nearly free electron metals with beyond-DFT methods}


\author{Subhasish Mandal}

\affiliation
{Department of Physics and Astronomy, West Virginia University, Morgantown, WV, USA}

\affiliation
	{Department of Physics and Astronomy, Rutgers University, Piscataway,  USA}

\author{Kristjan Haule}
\affiliation
{Department of Physics and Astronomy, Rutgers University, Piscataway,  USA}

\author{Karin M. Rabe}
\affiliation
{Department of Physics and Astronomy, Rutgers University, Piscataway,  USA}

\author{ David Vanderbilt}
\affiliation
{Department of Physics and Astronomy, Rutgers University, Piscataway,  USA}

\def\dvc#1{\textcolor{red}{[DV: #1]}}
\def\Red#1{\textcolor{red}{#1}}

\def\sm#1{\textcolor{cyan}{[SM: #1]}}
\def\smc#1{\textcolor{Brown}{#1}}

\begin{abstract}
{\footnotesize

\section{Abstract} 
For more than three decades, nearly free electron elemental metals have been a topic of debate because the computed bandwidths are significantly wider in the local density approximation to density-functional theory (DFT) than indicated by angle-resolved photoemission (ARPES) experiments.
Here, we systematically investigate this using first-principles calculations for alkali and alkaline-earth metals using DFT and various beyond-DFT methods such as meta-GGA, G$_0$W$_0$, hybrid functionals (YS-PBE0, B3LYP), and LDA+eDMFT. We find that the static non-local exchange, as partly included in the hybrid functionals, significantly increase the bandwidths even compared to LDA, while the G$_0$W$_0$ bands are only slightly narrower than in LDA. The agreement with the ARPES is best when the local approximation to the self-energy is used in the LDA+eDMFT method. We infer that even moderately correlated systems with partially occupied {\it s} orbitals, which were assumed to approximate the uniform electron gas, are very well described in terms of short-range dynamical correlations that are only local to an atom. 

}
\end{abstract}


\maketitle

\newpage

\section{Introduction}

Materials are often loosely categorized into weakly and strongly correlated systems depending on the strength of electron-electron correlation. Density functional theory (DFT) in the local density approximation (LDA), which maps each point in space inside a crystal to a uniform electron-gas problem through the electron density, is found to be fairly successful in the qualitative description of weakly correlated materials. However it is a ground-state theory, an accurate description of the excited-state properties even in moderately correlated materials lies outside the realm of conventional DFT. On the other hand, over the last few decades, there has been a longstanding effort to develop either perturbative, stochastic, or hybrid-functional approaches to understanding the strongly correlated and excited-state properties~\cite{onida_electronic_2002,DMFT2,DMFT3,kamal,PhysRevB.85.115104,jctc-mbj}. As a result, meta-GGAs, hybrid-functionals, the GW-approximation~\cite{HL,PhysRevB.88.125205}, and DFT+dynamical mean-field theory (DMFT) methods have become quite popular for understanding correlated solids. These methods are commonly referred to as ``beyond-DFT" methods. 

Beyond-DFT methods are often applied to open-shell {\it d}-electron systems, where the corrections to DFT are large and often qualitative, but very little is known about the applicability of these methods to solids that are only moderately correlated. In particular, the DFT+eDMFT method, which has gained wide popularity for describing the electronic structure of localized {\it d}- and {\it f}-electron systems, has so far mainly been applied to study strongly correlated systems~\cite{DMFT3,PhysRevLett.94.026404,Kunes:2008bh,PhysRevLett.109.156402,PhysRevLett.119.067004,PhysRevB.85.094505}, while its extension to the weak coupling limit is rarely discussed in the literature~\cite{PhysRevX.7.041067,PhysRevB.88.165119}. Similarly other ``beyond-DFT' methods such as meta-GGAs, hybrid functionals, and GW have mostly been applied to insulating materials~\cite{HL,PhysRevB.88.125205} with strong or moderate correlations, but it is unclear if these methods are equally successful when applied to weakly or moderately correlated metallic systems~\cite{PhysRevLett.88.016403,GW-metal}.  \\
 
In a recent work\cite{TMO1-SM}, we have systematically tested ``beyond-DFT" methods on insulating transition-metal monoxides, and showed that the hybrid functionals and eDMFT methods perform well, while other methods are inadequate in at least a couple of monoxides. Here, we use the same set of methods to investigate the alkali and alkaline-earth metals, whose electronic structure has been studied since the inception of quantum mechanics~\cite{Mahan_book}. They are often assumed to be the closest natural analogs to the uniform electron gas (UEG). 
DFT either with LDA or the generalized gradient approximation (GGA) is found to be fairly successful in the qualitative description of these metals, making them exemplary systems for weakly correlated electrons. 
However, the excited-state spectra and consequently the occupied bandwidths often quantitatively disagree with angle-resolved photoemission spectroscopy (ARPES) experiments. For example, ARPES studies by Lyo and Plummer~\cite{Na_ARPES1} indicate that the bandwidth of the first occupied band of sodium is substantially narrower than predicted by LDA. This has triggered a large amount of experimental~\cite{PhysRevB.30.5500, PhysRevB.33.3644,PhysRevB.41.8075,Sashin_2000,PhysRevB.41.3447,FS_Na} and theoretical~\cite{PRL_Mahan,PRB_Mahan,PhysRevLett.85.2410,theory1,PhysRevLett.83.2230,Na_QMC,doi:10.1143-JPSJ.68.3473,Na_DMFT,PRM_GW_DMFT,Kutepov_Na,Hybertsen_Na,PRB_Louie_2014,PhysRevLett.83.3250,PhysRevLett.85.2410,PhysRevLett.85.2411} work over the last three decades.

An earlier study using the G$_0$W$_0$ approximation~\cite{hedin_new_1965}, which truncates the self-energy to first order in the Green's function $G_0$ and the screened Coulomb interaction $W_0$, pointed to the importance of many-body effects and suggests that the lowest-order perturbative term may not be sufficient to describe the excitation spectrum in these materials~\cite{PRB_Mahan}. This motivated further GW studies to include higher-order perturbative corrections to the band limit~\cite{PhysRevLett.83.3250,PhysRevLett.85.2411,doi:10.1143-JPSJ.68.3473,PhysRevB.57.2108}~\cite{PhysRevB.86.035120,PhysRevB.100.054419}. Mohan and Sernelius~\cite{PRL_Mahan} and more recently Kutepov~\cite{Kutepov_Na} found that the inclusion of vertex corrections (GW$\Gamma$) modifies the GW self-energy, but the low-order vertex corrections to the self-energy and the dielectric function nearly cancel each other, resulting in a bandwidth renormalization not very different from the $G_0 W_0$ prediction. 

Recent work using a sophisticated unbiased reptation Monte Carlo method further indicates that the vertex corrections and self-consistency aspects of GW cancel to a large degree near the Fermi surface~\cite{PhysRevLett.107.110402}. This result also hints that the elemental metallic systems might be moderately correlated. An opposite trend of increasing the bandwidth compared to LDA is found in recent studies using self-consistent GW~\cite{PhysRevLett.83.3250,doi:10.1143-JPSJ.68.3473} and variational and fixed-node diffusion quantum Monte Carlo~\cite{PhysRevB.68.165103} techniques. Contrary to the indications from the GW approximation, Zhu and Overhauser~\cite{Overhouser} predicted that the spin fluctuations within a paramagnon pole model could account for the bandwidth reduction in Na, but a more recent study~\cite{PRB_Louie_2014} found that this effect is negligible. Interestingly, very recent ARPES results~\cite{Na_ARPES2} have ruled out the proposed~\cite{PRB_Louie_2014} strong coupling between the conduction electrons and spin fluctuations in Na and demands new theoretical insights. The entire situation remains unresolved, and the reason for the discrepancies between ARPES measurements and the theoretically predicted bandwidths in these simple metals remains one of the fundamental questions in condensed matter physics.

On a different tack, the narrowing of the ARPES spectra has alternatively been ascribed to final-state effects, which would require treating the outgoing electron as embedded in an interacting uniform electron gas inside the solid, rather than as a free electron leaving the solid~\cite{PhysRevLett.83.3250}. A similar approach was taken in Refs.~\cite{PRL0_Mahan} and ~\cite{PRB_Mahan}, but with the inclusion of surface effects. Such an interpretation was challenged in Ref.~\cite{PhysRevLett.85.2410} (see also \cite{PhysRevLett.85.2411}), as it would invalidate the accepted interpretation of the ARPES experiments as measuring the single-particle spectral function weighted by matrix-element effects~\cite{RevModPhys.75.473}. This would have far-reaching implications for the interpretation of all ARPES data to date.

To answer such questions, our computational study is carried out in the framework of a broader ongoing program involving the systematic performance and curation of electronic structure calculations using a range of methodologies and applied to a range of materials \cite{TMO1-SM,Choudhary2020}. The premise of this approach is that for a proper evaluation of the strengths and weaknesses of various first-principles methods applied to a given class of materials, a methodologically heterogeneous literature is not enough.  Rather, a set of calculations for different functionals performed on a consistent footing is essential. This is especially true in the context of high-throughput computation, where a desire for accuracy has to be carefully weighed against issues of consistency and modest computational load.  Moreover, the availability of these results in a materials database, such as the ``beyond-DFT'' component~\cite{bdft_database} of the JARVIS database~\cite{dft_database} used here, makes the comparisons between different functionals broadly available to the materials science community, providing a guide for future calculations on related systems.

Working in this context, we systematically apply several DFT and beyond-DFT methods to the elemental metallic systems from the first and second columns of the periodic table (Li to Cs and Be to Sr).  We resolve a controversy over the disagreement between theory and experiment for the occupied bandwidths of such systems, showing how they depend on the effects of local and non-local exchange and correlations.  We find that the band narrowing is surprisingly well described with non-perturbative dynamical correlations modeled as local to an atom rather than to a point in 3D space, emphasizing the importance of umklapp contributions to the electron self-energy at higher order in perturbation theory, beyond GW, seem to have a significant effect even in these systems.  In particular, in this letter we show that the elemental metals with partially occupied {\it s} orbitals, which are usually assumed to be nearly-free-electron metals, are in fact moderately correlated, thus forcing a reconsideration of long-held notions about these simple metals.

\begin{figure}
\includegraphics[width=250pt, angle=0]{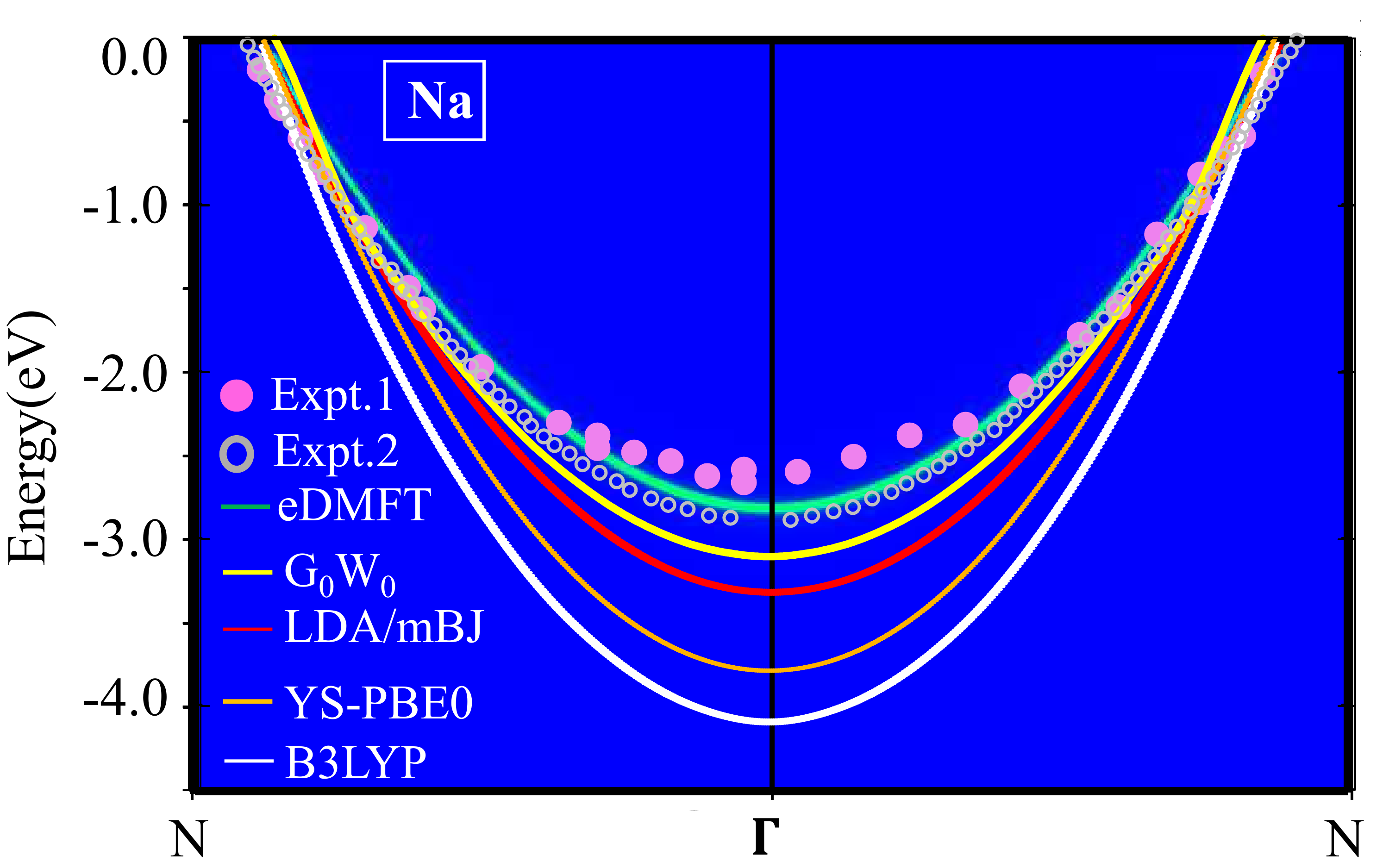}
\caption{ Band structure of elemental Na as computed in LDA, mBJ, G$_0$W$_0$, YS-PBE0, B3LYP, and eDMFT. 
 Dots in pink and grey indicate angle-resolved-photoemission(ARPES) data from earlier experiment by Lyo \& Plummer ~\cite{Na_ARPES1} and more recent experiment by Potorochin {\it et al.}~\cite{Na_ARPES2} respectively. }
\end{figure}

\begin{figure}
\includegraphics[width=240pt, angle=0]{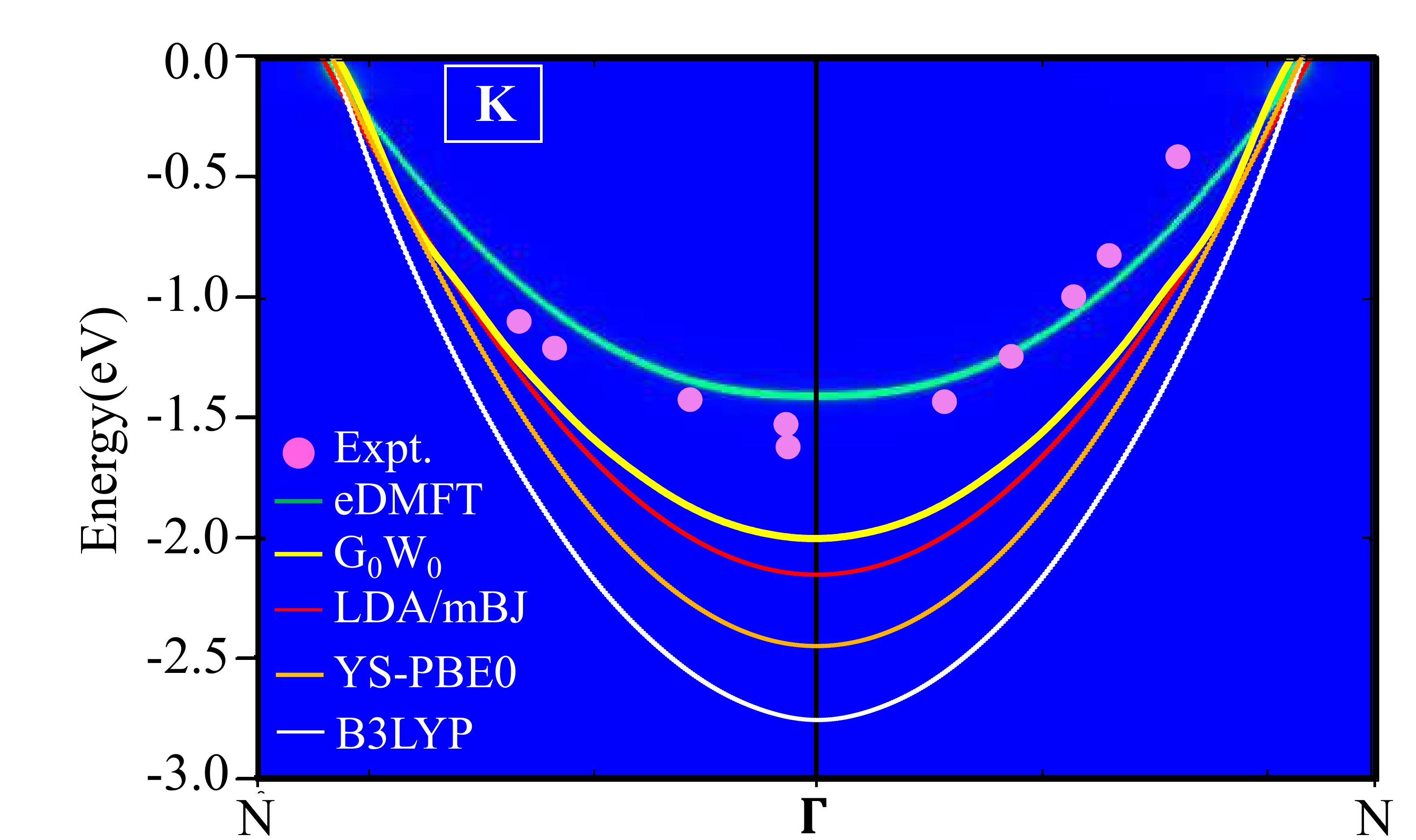}
\caption{ Band structure of elemental K as computed in LDA, mBJ, G$_0$W$_0$,YS-PBE0, B3LYP, and eDMFT. 
 Pink dots indicate angle-resolved-photoemission(ARPES) data from Ref.~\cite{PhysRevB.41.8075}.}
\end{figure}

\section{Results}

All computations are performed for the room-temperature experimental crystal structures obtained from the ICSD database. Most of the elemental metals studied here crystallize in the bcc structure at room temperature, except for Be and Mg which crystallize in hcp, and Ca and Sr in fcc. In the following we compare the electronic band structures using the above-mentioned methods with ARPES data, which are available for Na, K, and Mg. We describe each of these compounds in detail here, and direct the reader to the Supplementary Information for the others.

\subsection{Sodium} We first describe our results for Na, which has been most widely discussed in the literature as a prototypical UEG elemental metal. 
In Fig.~1, ARPES data are obtained from earlier experiment by Lyo \& Plummer ~\cite{Na_ARPES1} and shown as pink dots. Very recently an ARPES experiment for Na was repeated with more resolution by Potorochin {\it et al.}~\cite{Na_ARPES2} (shown as grey dots), indicating that the bandwidth is still underestimated as compared to LDA, and implying that a more advanced theoretical understanding is still needed.
The computed band structures in LDA, mBJ, B3LYP, G$_0$W$_0$, and eDMFT are shown in Fig.~1. We directly compare with the ARPES data (shown in pink and grey dots), which are adopted from Ref.~\cite{Na_ARPES1,Na_ARPES2}. Since the LDA and mBJ bands are found to be almost identical, we only show the LDA bands in the figure. While the bands are very similar to each other close to the Fermi surface near the $N$-point for all the methods, the differences between them become more evident near the $\Gamma$ point.

 Among these methods, the discrepancy between the B3LYP bands and the experiment is large. The agreement with ARPES for simple LDA is substantially better than for B3LYP. The screened hyrbid (YS-PBE0) performs slightly better than the unscreened one, but still substantially worse than LDA. We also compute the G$_0$W$_0$ bands, which is challenging for metallic systems due to numerical difficulties associated with the treatment of the Fermi surface singularity, often leading to less accurate results on the Matsubara axis, and consequently, extreme difficulty in the analytic continuation to real frequencies. This has been discussed in detail in Ref.~\cite{PyGW}, where the algorithm for the frequency convolution on the Matsubara axis was improved. This allows a stable analytic continuation of the imaginary-axis data via a Pade approximation, which is carried out for GW band structure calculations throughout the Brillouin zone (BZ).

It is evident in Fig.~1 that the G$_0$W$_0$ band (yellow) is very close to the LDA bands, narrowing only slightly relative to the LDA even at the $\Gamma$-point. The agreement between theory and earlier experiment is best for the eDMFT method; the spectral function (green) reproduces the ARPES data well throughout the BZ, except near the $\Gamma$ point where the agreement is slightly worse. This is again reconfirmed in the recent ARPES experiment~\cite{Na_ARPES2}, which was performed with more resolution than the earlier experiment by the Plummer group. The agreement now is better with LDA+eDMFT.

\subsection{Potassium}  We perform similar comparisons for bcc K in Fig.~2, where we plot the band structures as computed in LDA, mBJ, B3LYP, YS-PBE0, G$_0$W$_0$, and eDMFT. The ARPES data (shown in pink dots) are taken from Ref.~\cite{PhysRevB.41.8075}. For K, the ARPES data are not available for the entire BZ, extending only about halfway from $\Gamma$ to $N$. We again see that the bands obtained using various methods cross the Fermi energy at almost the same $k$ point, but the dispersion towards $\Gamma$ is quite different for the various methods. The B3LYP band disperses the most, reaching $-$2.76\,eV at $\Gamma$, while LDA and mBJ disperse only to $-$2.15\,eV. As was the case for Na, the LDA and mBJ bands are almost indistinguishable, so we display only the LDA in Fig.~2. The G$_0$W$_0$ bands narrow only slightly, reducing the bandwidth to 2.00\,eV from the LDA value of 2.15\,eV. The eDMFT narrows the bandwidth to 1.42\,eV at $\Gamma$, as compared with the experimental value of 1.6\,eV. (Slightly away from $\Gamma$, the ARPES data are asymmetric which could be due to inaccuracy in the experimental data~\cite{PhysRevB.41.8075}. As a result, the eDMFT band crosses the ARPES band only on the right half of the figure.  A similar artifact appears for Na in Fig.~1.)

\subsection{Magnesium}
 Unlike most of the elemental metals, Mg crystallizes in the hcp instead of the bcc structure at room temperature and has multiple occupied bands.  We display the bands in Fig.~3. The ARPES data (pink dots, taken from Ref.~\cite{PhysRevB.33.3644}) indicate two bands within this energy range, one with small and one with large dispersion from the Fermi energy to $\Gamma$. In particular, one band disperses from $\Gamma$ at $-$0.82\,eV up to the Fermi energy $(E=0)$, and another disperses from $-$1.85\,eV down to $-$6.15\,eV along $\Gamma$--A--$\Gamma$.
 Here we also find a similar trend in the computed bands using various methods as noticed for Na and K. 
 The B3LYP bands disperse the most and lie furthest from the experimental bands. Once again, the LDA and mBJ bands are both very similar. The first G$_0$W$_0$ band disperses to $-$1.29\,eV at $\Gamma$, while in eDMFT it is $-$0.82\,eV and in the experiment, it is $-$0.9\,eV. The second occupied band crosses the $\Gamma$ point at $-$6.66, $-$6.18, and $-$6.15\,eV in G$_0$W$_0$, eDMFT, and experiment, respectively. We again find the best overall agreement with ARPES for the eDMFT spectral function. We find similar trends in the band structures of other elemental metals in this family as well, as discussed in the Supplement. The sharp spectral function in eDMFT, which is common in Figs. 1-3 reflects that electron-electron scattering is weak in these metals.

\begin{center}
\begin{table*}

\begin{tabular}{ l  r r  r  r r r  r}

\hline
\hline
{Compound \vspace{0.2in}} & {LDA} & {mBJ} & {B3LYP} & {YS-PBE0}& {eDMFT} &{G$_0$W$_0$}  & {Expt} \\
\hline

\vspace{0.1in}
\multirow{1}{*}{Li}& 3.46 & 3.30 & 4.22 & 4.41 & 2.60 & 3.39 &  \\

\multirow{2}{*}{Be} & 4.28 & 4.25 & 5.05 & 4.89 & 4.41 & 4.48 & 4.8~\cite{PhysRevB.30.5500}\\
\vspace{0.1in}
&  11.03 & 10.90 &  12.73 & 12.74 &10.12 & 11.37 & 11.1~\cite{PhysRevB.30.5500}\\

\vspace{0.1in}
\multirow{1}{*}{Na}  & 3.30 & 3.29 & 4.09 & 3.79 & 2.84 & 3.15 & 2.65~\cite{Na_ARPES1} \& 2.78~\cite{Na_ARPES2} \\

\multirow{3}{*}{Mg} & 1.31 & 1.35 &  1.63 & 1.37 & 0.82 & 1.29 &  0.9~\cite{PhysRevB.33.3644} \\

& 1.65 &  1.61 &2.07 & 1.95 &  1.85& 1.68 & 1.70~\cite{PhysRevB.33.3644} \\
\vspace{0.1in}
 &  6.89 &  6.89 & 8.09 & 7.89 &6.18 &  6.66 & 6.15~\cite{PhysRevB.33.3644} \\

\vspace{0.1in}
\multirow{1}{*}{K} & 2.15& 	2.13&2.76 & 2.46 &1.42 &2.00 & 1.60~\cite{PhysRevB.41.8075}\\
\vspace{0.1in}
\multirow{1}{*}{Ca} & 3.98 &3.82& 4.88 & 4.65	& 3.24 &	3.79 &	3.30~\cite{Sashin_2000}\\
\vspace{0.1in}
\multirow{1}{*}{Rb} & 1.99 & 1.96& 2.53 & 2.23  & 1.81  & 1.86  &    \\
\vspace{0.1in}
\multirow{1}{*}{Sr} & 3.7	& 3.48& 4.46 &4.21	&3.05	&3.39 &  \\

\multirow{1}{*}{Cs} & 2.15 & 1.95 & 2.60 & 2.37 & 1.70 & 2.00 & \\
\hline
\hline
\end{tabular}
\caption{Bandwidth (in eV) of occupied bands for various elemental metals as computed in various beyond-DFT approaches and their comparison with available experiments with 0-12 eV.}
\label{tab:compare}

\end{table*}
\end{center}

\begin{figure}
\includegraphics[width=240pt, angle=0]{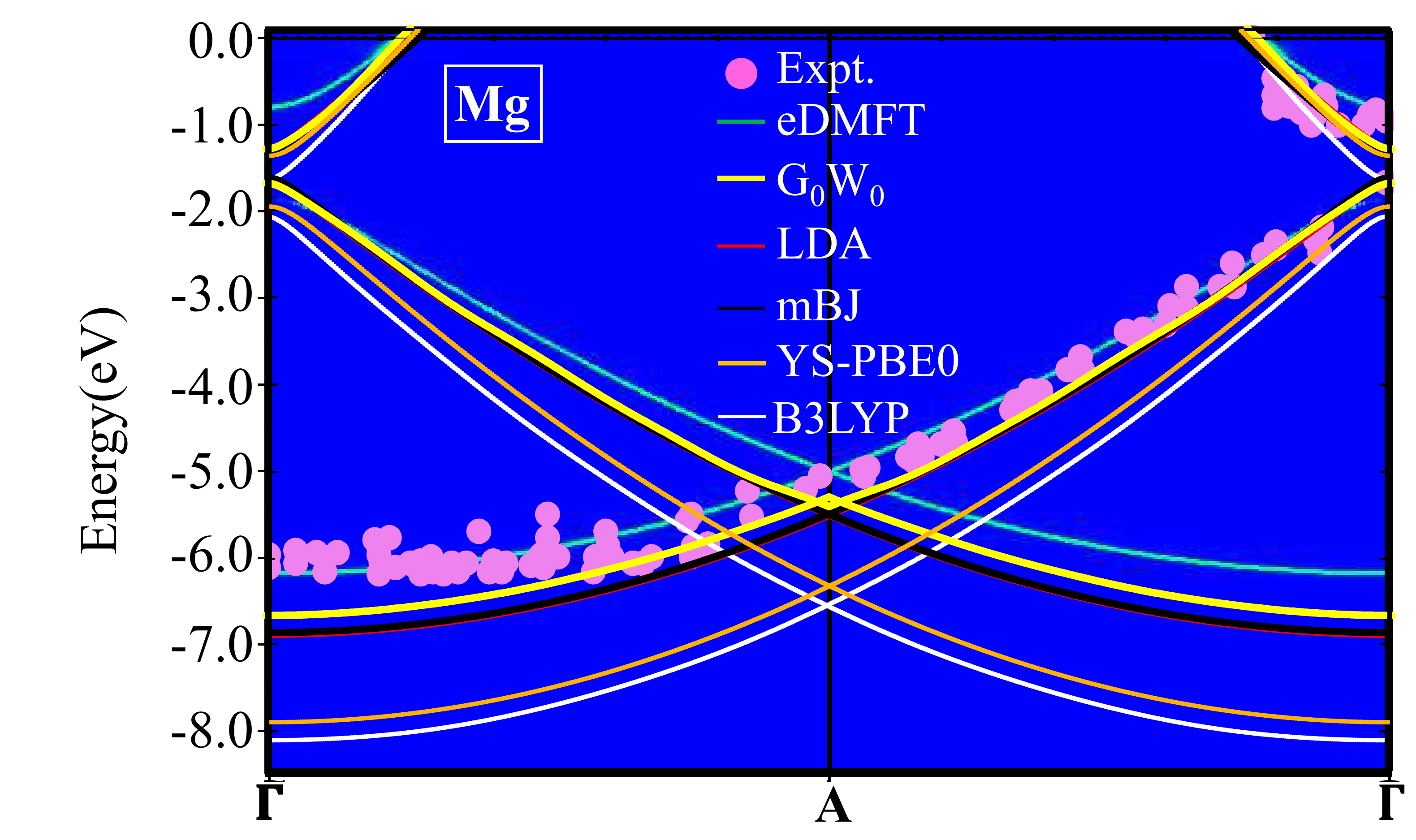}
\caption{ Band structure of elemental Mg as computed in LDA, mBJ, G$_0$W$_0$, YS-PBE0, B3LYP, and eDMFT. Pink dots indicate angle-resolved-photoemission(ARPES) data which are adopted from Ref\cite{PhysRevB.33.3644}.
}
\end{figure}

\subsection{Bandwidth of elemental metals} To quantify the observed trends in alkali and alkaline-earth metals, we compare the bandwidths of the occupied bands, defined as the depth of the energy at $\Gamma$ below $E_F$, as computed using LDA, mBJ, B3LYP, YS-PBE0, G$_0$W$_0$, and eDMFT, in Table I. We describe the bandwidths within a 0-9 eV window in Fig. 4(a) and the relative error from the experimental value in Fig. 4(b).

We also compare them with available experiments from ARPES as well as other photoemission spectroscopies. As the Fermi surface is almost exactly spherical in these compounds, and the band structure is close to a renormalized free-electron solution in the proper periodic potential, the most relevant number here is the bandwidth.
LDA overestimates the bandwidths across the family as expected, except for Be. 
The relative error in LDA as compared to experiment goes as high as 56\% (Fig. 4b).   
The hybrid functional, which corrects some of the self-interaction error by incorporating a fraction of exact exchange, correcting the non-local part of the self-energy, has been found to significantly improve the descriptions of many {\it d}-electron systems~\cite{PhysRevB.74.155108,doi:10.1063/1.1564060,PBE0,LYP,TMO1-SM}, in particular Mott insulators. Here we find that the static non-local corrections included either in B3LYP or YS-PBE0 have the opposite effect, worsening the agreement with ARPES as they significantly overestimate the bandwidth for all the systems studied here, with relative error, which can reach as high as 80\% (Fig. 4b). This clearly shows that the hybrid functionals do not improve on LDA for metallic systems~\cite{doi:10.1063/1.2747249}. mBJ performs better than B3LYP and gives very similar bandwidths as LDA, except for heavy elements such as Sr and Cs, where mBJ performs better than LDA. Next, we notice that the bandwidths as computed in the G$_0$W$_0$ method are in better agreement than LDA or mBJ, but are still substantially wider than those measured by ARPES. In  G$_0$W$_0$, the relative error varies from 1 to 43\% across the compounds studied here. Various implementations of the GW approximation, with and without vertex corrections, give the bandwidth for Na in the range of 2.5-3.2\,eV~\cite{Hybertsen_Na,Kutepov_method}. Here we obtain 3.15\,eV, quite similar to recently reported self-consistent quasiparticle GW values of 3.17\,eV~\cite{Kutepov_method}. This result indicates that  single-shot GW or G$_0$W$_0$ perform similar to that of self-consistent quasiparticle GW and explains that for the elemental metals self-consistency may not be necessary. The improvement of GW as compared to static hybrid methods, which incorporate some Hartree-Fock self-energy, shows that the dynamical nature of the correlations, with proper description of screening, is certainly an important factor in the band narrowing in Na and other alkali metals.\\

\begin{figure}
\includegraphics[width=240pt, angle=0]{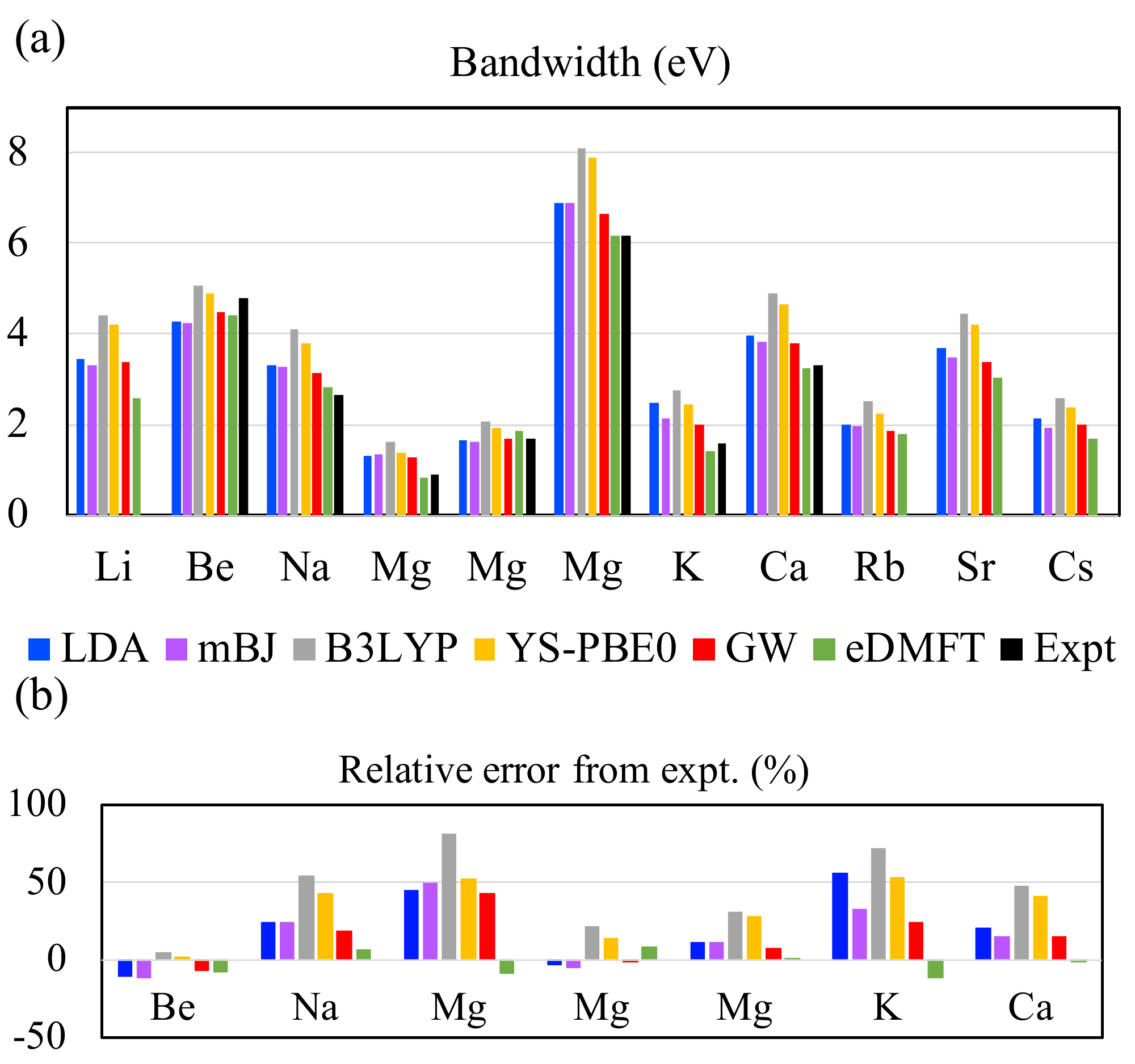}
\caption{ Comparison of bandwidths for alkali and alkaline-earth metals: (a) bandwidth values (in eV) as obtained from various beyond-DFT approaches and (b) their relative error from the experiment. }
\end{figure}

Finally, the bandwidth narrowing is even stronger in eDMFT due to the dynamic correlations incorporated in the strong frequency dependence of the self-energy, which further improves the agreement with ARPES experiments. This indicates that the umklapp processes at higher order in perturbation theory, beyond GW, seem to have a significant effect even in these relatively weakly interacting solids. The relative error in eDMFT varies between 0.5 to 11\% (Fig. 4b) across the compounds studied here. While within DFT+$U$, the static analog of LDA+eDMFT, the bandwidth is insensitive to the value of $U$ (not shown here), in the LDA+eDMFT results here, the bandwidth does depend on the local Coulomb interaction $U$ (see the  Supplementary Information ). 

\section{Discussion}

Here, we have systematically studied the electronic structure of elemental metals from the first and second columns of the periodic table using LDA and various beyond-DFT methods, such as the mBJ meta-GGA and B3LYP and YS-PBE0 hybrid functionals, GW, and LDA+eDMFT (eDMFT), and compared with ARPES experiments where available. We have found that B3LYP significantly overestimates the bandwidths, even compared to LDA and mBJ, which are close to each other. G$_0$W$_0$ reduces the bandwidths further, but still not sufficiently, while eDMFT narrows the bands the most, producing spectral functions that match rather well with the ARPES experiments. These trends were found to follow almost uniformly over the elemental metallic family of compounds studied here.

It is well known that the dispersion relation for the Hartree-Fock (HF) equations applied to the ground state of the uniform electron gas or the jellium introduces a logarithmic divergence of the derivative of the band energy at the Fermi level~\cite{Ashcroft,doi:10.1063/1.4909519}. This gives an infinite velocity and zero density of states at the Fermi energy for the UEG. Many-body correlation effects screen the Coulomb potential and eliminate the divergence. 

Since hybrid functionals contain a portion of Hartree-Fock exchange, it is expected to perform poorly in the context of metals, which is evident from our computations for two different hybrid functionals, namely B3LYP and YS-PBE0 (similar to HSE06).  In particular, since B3LYP contains the bare (unscreened) exact-exchange potential, the comparison with experiment is worse than with YS-PBE0 or HSE06 where the interaction is screened. The fact that YS-PBE0 improves only slightly over B3LYP, suggests that screening in metals is far stronger than assumed in the standard hybrid functionals, and from comparing to other methods, we can conclude that the dynamic umklapp processes rather than static non-local (to an atom) effects dominate.

On the other hand, meta-GGAs such as the mBJ method, which is again popular for insulating or semiconducting materials, can be seen as a kind of ``hybrid" potential whose amount of ``exact exchange" is controlled by a parameter that controls the separation between short and long-range exchange in the screened hybrid functional~\cite{mbj}. As a result, non-locality in this method is restricted, and it performs much better than hybrid functionals for the simple metals. The performance of DFT was known to improve if we go up in the famous ``Jacob's ladder” for the exchange-correlation energy~\cite{doi:10.1063/1.1390175}. Here we find an opposite trend for simple metals, where meta-GGA performs much better than hybrid functionals (screened or unscreened), and the difference between LDA and mBJ is not significant. It is worth mentioning that the meta-GGAs, in particular the mBJ method, which recovers the LDA bandwidth for metallic system, is computationally as expensive as LDA and considerably cheaper than either hybrid functionals or GW method. 
 Whether this method is suitable for metals was not clear, as it overestimates magnetic moments in ferromagnetic metals~\cite{PhysRevB.83.195134}. Here we demonstrate that the mBJ method is applicable to non-magnetic simple metallic systems, especially for the heavy alkali or alkaline-metals where it performs better than LDA. 
\\

In LDA, mBJ, or hybrid functionals the correlation effect is static. It is well known that the dynamic nature of correlation incorporated in GW, which also contains non-local self-energy, can overcome the barrier of failure of conventional-DFT methods. Here we find that the bandwidths computed in GW can improve over LDA, but not to the extent needed to reach agreement with the experimental bandwidths.

It is intriguing to understand why the bands in LDA and G$_0$W$_0$ have almost identical slopes near the Fermi energy for almost all elemental metals. As obtained from Landau Fermi liquid theory, the expressions for the difference in band mass in LDA and G$_0$W$_0$, are~\cite{Mahan_book,KH-DMC} 
\begin{eqnarray}    
\frac{1}{m_{DFT}}-\frac{1}{m_{GW}} &=&  \left[\frac{\partial V_{xc}}{\partial k} -Z_k\frac{\partial \Sigma}{\partial k}\right]\frac{1}{k_F} + \frac{1-Z_k}{m}
\end{eqnarray}
where $V_{xc}$, $\Sigma$ and $Z_k$ are the exchange-correlation potential in DFT, GW self-energy, and quasiparticle renormalization amplitude in GW, respectively. $m$ is the bare band mass, and $k_F$ is the Fermi wavevector. Since most of the metals studied here have very similar bandwidths in LDA and GW, we can infer from the above equations that the changes coming from $Z$ and  $\partial\Sigma/\partial k$, largely cancel each other to give similar slopes in LDA and GW near the Fermi energy. 
Indeed, $V_{xc}$ is local in LDA, hence we can conclude that $\partial\Sigma/{\partial k}\approx(1/Z_k-1) k_F/m$ is positive and it reduces the mass, while $Z_k<1$ always increases the mass.

In LDA+eDMFT the mass is renormalized only by $Z$, as the self-energy has no momentum dependence. Hence, the cancellation between the momentum and frequency derivative does not occur, and the band narrowing is stronger. Given the much better agreement of eDMFT with the ARPES, this would suggest that the momentum dependence of the self-energy in this moderately correlated regime might be overestimated by single-shot {G$_0$W$_0$}, and that a better non-perturbative treatment of correlation effects should increase the frequency dependence and reduce the momentum dependence of the self-energy.  Perhaps the self-consistent GW$\Gamma$ calculation with vertex corrections can solve this problem. Based on our study, however, we cannot exclude the possibility that the conventional interpretation of the ARPES experiments is invalid, and that the experimental narrowing of the bands is an artifact of final-state interactions~\cite{PhysRevLett.83.3250} or surface effects~\cite{PRL0_Mahan,PRB_Mahan}, opening up new possibilities and challenges for ARPES experiments in future~\cite{Na_ARPES2}. A recent ARPES measurement, which tries to eliminate some of these effects in the experiment, agrees with LDA+eDMFT more strongly than the earlier experiment and reconfirms which predictive method is powerful for simple metals. Our work also shows that the umklapp contributions to the electron self-energy at higher order in perturbation theory (beyond GW) are important and have a significant effect in alkali and alkaline-earth metals.

Our study suggests that the elemental metals with partially occupied {\it s} orbitals are better described in terms of short-range dynamical correlations (local to an atom rather than to a point in 3D space) with predominantly large momentum umklapp contributions than in the weakly interacting perturbative picture, or in non-local hybrid schemes that incorporate purely local correlations and non-local Hartree-Fock exchange. Such a short-range dynamical description was already very successful for strongly correlated systems with partially filled $d$ or $f$ bands. Here, we conclude that it is also remarkably successful in describing moderately correlated simple metals as well.

 \section{Methods} 

 In this work we have used the full potential linear augmented plane wave (LAPW) method, as implemented in the WIEN2k~\cite{WIEN2k} software and its extensions. The following DFT functionals in WIEN2k software are used: LDA/GGA, Becke-Johnson (mBJ) potential~\cite{mbj} for meta-GGA, and B3LYP~\cite{doi:10.1063/1.464913,PhysRevB.74.155108} for hybrid functionals. The GW extensions of WIEN2k was implemented in PyGW software~\cite{PyGW} following the Hedin's GW formalism, and embedded dynamical mean field method (eDMFT) functional is implemented in LDA+eDMFT software~\cite{eDMFT2010}. For G$_0$W$_0$,  mBJ and B3LYP hybrid functionals (B3LYP and YS-PBE0) we construct the initial wavefunction and eigenvalues with the LDA functional as a starting point, and then fully self-consistent calculations were achieved in all but G$_0$W$_0$, which requires a single shot beyond LDA.
 For YS-PBE0, which is very similar to HSE06 functional, we use a screening parameter of 0.165. In B3LYP there is no screening parameter. The standard DFT computations with the hybrid functionals were first performed self-consistently, and then the bandstructures were obtained non self-consistently. The interpolation methods, as implemented in WIEN2k, assume smooth energy bands in hybrid functionals to avoid any divergence. Thus, the anomalous band dispersion for B3LYP is not seen in the bandstructures. \\ 

\textbf{LDA+eDMFT:} In this work, we use the embedded implementation of the all-electron DFT+DMFT method, where the self-energy is approximated by a quantity local to the atom in the unit cell. 
The non-perturbative form of such self-energy is obtained by solving the quantum impurity problem in the presence of a self-consistent mean-field environment. This work uses the fully self-consistent DFT-DMFT implementation developed at Rutgers by one of the co-authors. The quantum impurity method is solved by the continuous-time quantum Monte Carlo method~\cite{haule2007}.
The self-energy for the s-orbital is embedded by the Dyson equation into the Hilbert-space of all electrons in the solid using projection/embedding technique~\cite{eDMFT2010}.
This projection step connects the atomic degrees of freedom with the continuum degrees of freedom of solid, and leads to causal DMFT equations, which can capture spectral weight of the all electrons in the solid. Unlike DFT functional, the DFT+DMFT functional is based on the Baym-Kadanoff functional, which delivers not only the ground-state, but also the excited state properties of solids. The eDMFT implementation allows one to write down the total energy in terms of the stationary form of the energy functional.
Unlike many other DFT+DMFT implementations, we do not downfold to Wannier orbitals, but rather use projectors to the very localized orbitals contained within the muffin-tin spheres. 
The purpose of the projection, which is done on correlated orbitals within Fermi-energy $\pm$ 10 eV, is to extract the local Green’s function from the full Green’s function. For the case of simple metals, we have considered {\it s}-orbitals to be correlated. We have also tested this with including the {\it p}-orbitals.
In this work LDA+embedded-DMFT(LDA+eDMFT) method~\cite{eDMFT2010} means that we used the combination of DMFT and the LDA functional in the LAPW basis set as implemented in WIEN2k~\cite{WIEN2k}.
Using the {\it exact} double counting between LDA and eDMFT~\cite{ExactDC}, we obtain the self-energy on Matsubara frequency, which is then analytically continued with the maximum entropy method from the imaginary to the real axis, continuing the local cumulant function, to obtain the partial density of states and the spectral functions. In eDMFT, where all such higher-order Feynman diagrams are explicitly calculated by the impurity solver, the amount of the screening by the degrees of freedom not included in the method, is substantially reduced, and consequently the values of $U$ are larger and more universal across similar set of material, and are quite successfully predicted by the self-consistent constrained method, similar to the constrained LDA, but using the LDA+eDMFT functional to evaluate total energies. A fine k-point mesh of at least 15$\times$ 15 $\times$ 15 k-points in Monkhorst-Pack k-point grid and a total of 20 million Monte Carlo steps for each iteration are used for the elemental metals at $T=300\,$K. To avoid fine tuning parameters, the Coulomb interaction $U$ and Hund's coupling $J_{H}$ are fixed at 5.0 eV and 0.3 eV respectively. 
 To estimate these values of the Coulomb repulsion, we used the constrained-DMFT method, in which one computes the total energy in a supercell for $N-1$, $N$ and $N+1$ constrained electrons, where $N$ is the number of constrained electrons in the localised orbital. Then we take the difference of energies $E(N+1)-E(N)$ and $E(N)-E(N-1)$ to compute the value of $U$, just like in constrained LDA.
The computed values of $U$ for the presented metallic systems within the self-consistent constrained-eDMFT method are close to $U=5\,$eV, a value which we adopt for all the compounds studied here. Thus, $U$ is not treated as a tuning parameter. The computed values of $U$ do not vary much among these metallic compounds. We nevertheless studied the dependence of the spectral function on $U$ as shown in Fig. S1. for Na. Surprisingly, a similar value of $U$ was obtained for strongly correlated metals like FeSe, and using U$=$5 eV, the spectral functions matched remarkably well with experiments in various Fe-chalcogenides and pnictide compounds~\cite{haule3}. \\

\textbf{GW:} We perform single-shot GW (G$_0$W$_0$) using PyGW software package~\cite{PyGW}, an all-electron GW implementation, where GW self-energy is computed within the all-electron LAPW basis of WIEN2K. In this newly developed GW code, special attention is paid to the metallic systems and proper treatment of deep laying core states.
Prior to GW calculation, we perform LDA calculation to obtain the input single-particle Kohn-Sham wavefunctions($\psi_{\vk,i}$) as well as the eigenvalues ($\varepsilon_\vk^0$ ) in the LAPW basis. 
In the LAPW method, the space is partitioned into non-overlapping atom-centered muffin tin (MT) spheres and the
interstitial region (IR). We included core electrons in our GW computations. The radius of MT for the elemental metals were considered to be 2.5 Bohr. A 12 Ry cut off energy for the plane wave was used and R-MT*K-MAX was fixed at 7 atomic unit. The number of core states and number of the empty bands used for construction of polarizability matrix are described in Table II in the Supplementary Information for all the compounds studied here.
In the G$_0$W$_0$ method, we obtain the screened interaction $W_0$ from the Polarization $P$, which is computed by the convolution of the two single-particle Kohn-Sham Green's functions $G_0=1/(\omega+\mu-\varepsilon_\vk^0$) in the form $P=-iG_0*G_0$. Here $\mu$ is the chemical potential and $\omega$ is the frequency. Polarization $P$ and the screened interaction $W$ are expressed in the product basis~\cite{ProductBasis0}, which is further rotated to diagonalize the Coulomb interaction matrix. This product basis allows one to express $P$, $W$ and the Coulomb interaction in the matrix form.  The product basis~\cite{ProductBasis0} is an orthogonal over-complete basis that can faithfully represent products of two Kohn-Sham orbitals, $\xi_\alpha^\vq(\vr)$. Here $\vr$ stands for the real space vector, and $\vq$ is momentum in the first Brillouin zone. In this implementation within the LAPW basis,  the product basis functions $\ket{\xi_\alpha}$ are not only orthonormal in the muffin tin part, they are also made orthonormal in the interstitial part, unlike in most other implementations~\cite{ProductBasis0,Bluegel,Gap2}. More technical details are available in  Ref~\cite{PyGW}.

Once such product basis $\xi_\alpha^\vq(\vr)$  is constructed, we compute the matrix elements between two Kohn-Sham orbitals and this basis functions: $M_{\alpha,ij} (\vk,\vq)\equiv \braket{\xi^\vq_\alpha| \psi_{\vk,i}\psi^*_{\vk-\vq,j}}$ as well as  M$_{\beta,i'j'}^*(\vk',\vq)\equiv$ $\braket{\psi_{\vk',i'}\psi_{\vk'-\vq,j'}^*|\xi^\vq_{\beta}}$, where $\psi_{\vk,i}$, $\psi_{\vk'-\vq,j'}$ are incoming, and $\psi^*_{\vk-\vq,j}$, $\psi_{\vk',i'}^*$ are outgoing electrons. Similarly, we compute the matrix elements of the Coulomb repulsion in this basis $v_{\alpha\beta} (\vq)=\braket{\xi^\vq_\alpha|V_C(\vq)|\xi^\vq_\beta}$, and then compute the square root of the Coulomb repulsion in its eigenbasis $\sqrt{ v(\vq)}_{\alpha,\beta}= T_{\alpha,l}\sqrt{v_l} \, T^\dagger_{l,\beta}$, where $v_l$ are eigenvalues and $T_{\alpha,l}$ are eigenvectors of the Coulomb repulsion. 

The dielectric function in matrix form is computed as $\varepsilon = 1 - \sqrt{V_C} P \sqrt{V_C}$. The advantage of the product basis is elaborately described in Ref.~\cite{PyGW,Bluegel}. Once the dielectric matrix $\varepsilon$ is calculated, we invert it in this eigenbasis of the Coulomb repulsion to compute the screened interaction $W$. The dynamical correlation self-energy within GW approximation is then obtained by convolution of $W$ and the single-particle Kohn-Sham Green's function $G_0$. In PyGW code, the dynamic part of the screened interaction is computed on the Matsubara axis, which performs much better than classical plasmon-pole method~\cite{EFPP}. For the analytic continuation to the real frequency axis we  used standard Pade method~\cite{Pade1,Pade0}, which is accurate at low to moderate frequencies, provided we have very accurate imaginary axis data. This was achieved in our calculations up to 5 eV, which is sufficient to plot reliable band structures of these metallic materials. The Pade approximation is forced to go exactly through all Matsubara frequencies calculated on logarithmic mesh (between 32-64), hence the number of poles in such analytic function is large (between 30-62) since for metals, a few pole approximation in Pade-type fitting is usually not sufficient. We then carefully converge our results using a very fine k-point mesh as the convergence for metallic systems can be difficult in GW method. Here, even though we use tetrahedron analytic integration over momentum points, we find that a large number of momentum points is necessary for convergence. For example, although the $4\times 4 \times 4$ grid gives approximate spectra not too different from LDA, the convergence is reached only at $16\times 16 \times 16$ momentum mesh.  We have used the interpolation method from Ref~\cite{Pickett_method,Pickett_method0} for computing spectra along the high-symmetry lines. The input data at momentum points in the calculation are of high quality, but the smooth interpolation at other momentum points is not very straightforward in the delocalized basis sets, especially for interstitial charge and such interpolation is not extremely precise at points where cusps occur in the band-structure, and sometimes leads to small computational artifacts near Fermi surface singularity. 
For this reason, we notice the GW-bands have some dispersion anomalies close to the Fermi energy in some of the computed spectra. 
The Fourier transformation of the self-energy from momentum space to the real space, and its evaluation at the generic momentum point along the momentum path, will improve on the current simple interpolating scheme. More sophisticated and more accurate Wannier interpolation might also ameliorate this issue and is beyond the scope of this work. For more technical details about methodology, implementation of GW in LAPW basis as well as the K-point convergence, we refer the reader to Ref.~\cite{PyGW}. \\

{\it Note added in proof:} While preparing this manuscript we became aware of a recent ARPES experiment~\cite{Na_ARPES2}, that reconfirms our prediction with LDA+eDMFT for Na.

\section{Data availability}
The data that support the findings of this study are available from the corresponding authors upon request. In the future, the data will be available in the form of an open-source database that is currently under construction.

\section{Author Contributions}
S.M carried out the calculations. All authors discussed the results and co-wrote the paper.

\section{Competing interests} The authors declare no competing interests.

\section{Code availability}
We have used the WIEN2K~\cite{WIEN2k} DFT+eDMFT, and PyGW codes to generate the data. WIEN2K is a commercially available software package. LDA+eDMFT and PyGW are freely available and can be obtained from \url{http://hauleweb.rutgers.edu/tutorials} and \url{https://github.com/ru-ccmt/PyGW} respectively.

\section{Acknowledgments } 
 This research was funded by NSF DMREF DMR-1629059 and NSF DMREF DMR-1629346. The computations were performed at  the Frontera supercomputer at the Texas Advanced Computing Center (TACC) at The University of Texas at Austin, which is supported by National Science Foundation grant number OAC-1818253 and at the Extreme Science and Engineering Discovery Environment (XSEDE), which is supported by National Science Foundation grant number ACI-1548562.


\bibliography{SM-bib}

\vspace{4 in} 

\pagebreak
\renewcommand{\thefigure}{S\arabic{figure}}

\setcounter{figure}{0}

\renewcommand{\thetable}{S\arabic{table}}

\setcounter{table}{0}

\newpage
\section{Supplemental Material}

\section{Supplementary Methods}



\vspace{-0.1 in} 
\begin{table}
\begin{tabular}{ c   c  }

\hline
\hline
{Compound \vspace{0.2in}} & {ICSD-ID} \\
\hline
\multirow{1}{*}{Li}& 44367\\	
\multirow{1}{*}{Be}& 1425 \\	
\multirow{1}{*}{Na} & 196972	\\
\multirow{1}{*}{Mg} & 76748	\\
\multirow{1}{*}{K}& 44670	\\
\multirow{1}{*}{Ca}& 44348 \\	
\multirow{1}{*}{Rb}& 44869	\\
\multirow{1}{*}{Sr} & 76162	\\
\multirow{1}{*}{Cs} & 42662 \\	
\multirow{1}{*}{Ba}& 96587\\

\hline
\hline
\end{tabular}
\caption{ICSD-IDs for compounds studied here.}
\label{tab:icsd}
\end{table}

\textbf{Crystal Structures:} The experimental crystal structures are obtained from the ICSD-database. The ICSD numbers are given in Table ~\ref{tab:icsd}.

\textbf{GW-details:} The number of unoccupied bands and the number of core states included in the GW calculations are described in Table ~\ref{tab:GW}

\begin{table}
\begin{center}
\begin{tabular}{ c   c  c }

\hline
\hline
{Compound \vspace{0.2in}} & {\#unoccupied bands} & {\#core state} \\
\hline
\multirow{1}{*}{Li}& 40 & 0\\	
\multirow{1}{*}{Be}& 62 & 2\\	
\multirow{1}{*}{Na} & 88 & 1\\
\multirow{1}{*}{Mg} & 126 & 2\\
\multirow{1}{*}{K}& 188 & 8\\
\multirow{1}{*}{Ca}& 109 & 8\\	
\multirow{1}{*}{Rb}& 225 & 25	\\
\multirow{1}{*}{Sr} & 143 & 25\\
\multirow{1}{*}{Cs} & 177 & 32\\	
\multirow{1}{*}{Ba}& 225& 25 \\

\hline
\hline
\end{tabular}
\caption{Number of unoccupied bands and core states considered for constructing Polarizability matrix in GW calculations.}
\label{tab:GW}
\end{center}
\end{table}

\section{Supplementary Discussion}
\begin{figure*}
\includegraphics[width=480pt, angle=0]{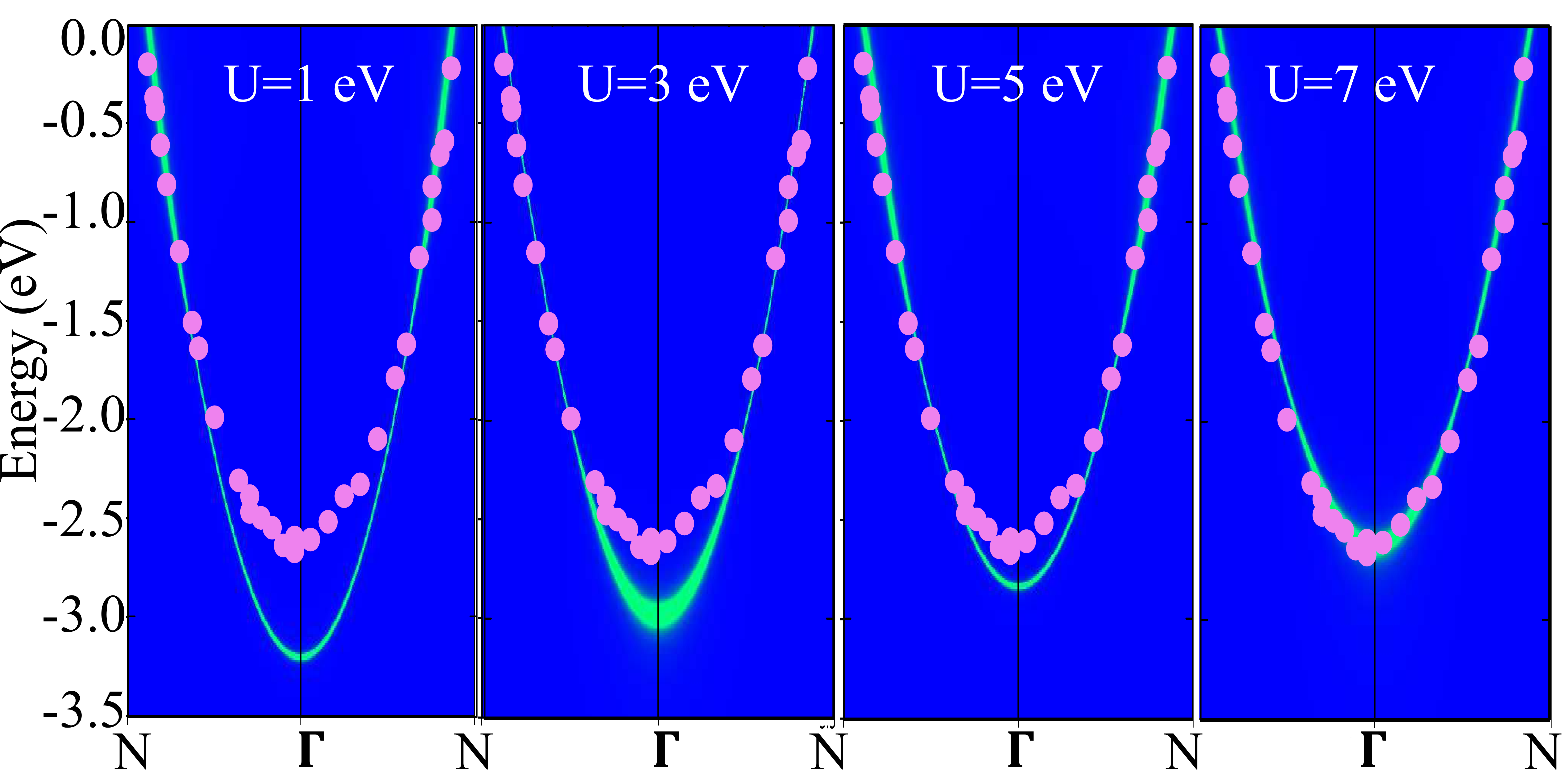}
\caption{The dependence of Coulomb $U$ in eDMFT spectral function for elemental Na. Pink dots indicate angle-resolved-photoemission (ARPES) data from Lyo and Plummer, Phys. Rev. Lett. 60, 1558 (1988).}
\end{figure*}

First, in Supplementary Figure 1, we show the Coulomb $U$ dependence of eDMFT spectral function for elemental Na. The bandwidth shows a strong dependence on the Coulomb $U$. Pink dots in Fig. S1 indicate angle-resolved-photoemission (ARPES) data from Ref.~\cite{Na_ARPES1}.\\

The imaginary part of the eDMFT self-energy gives the scattering rate of electrons and also gives additional microscopic insights into the correlated electronic structure evolution. 
In Supplementary Figure 2, we show the frequency dependence of the imaginary part of the self-energy for s-orbital in Mg, Ca, Sr, and Rb.
\begin{figure*}
\includegraphics[width=440pt, angle=0]{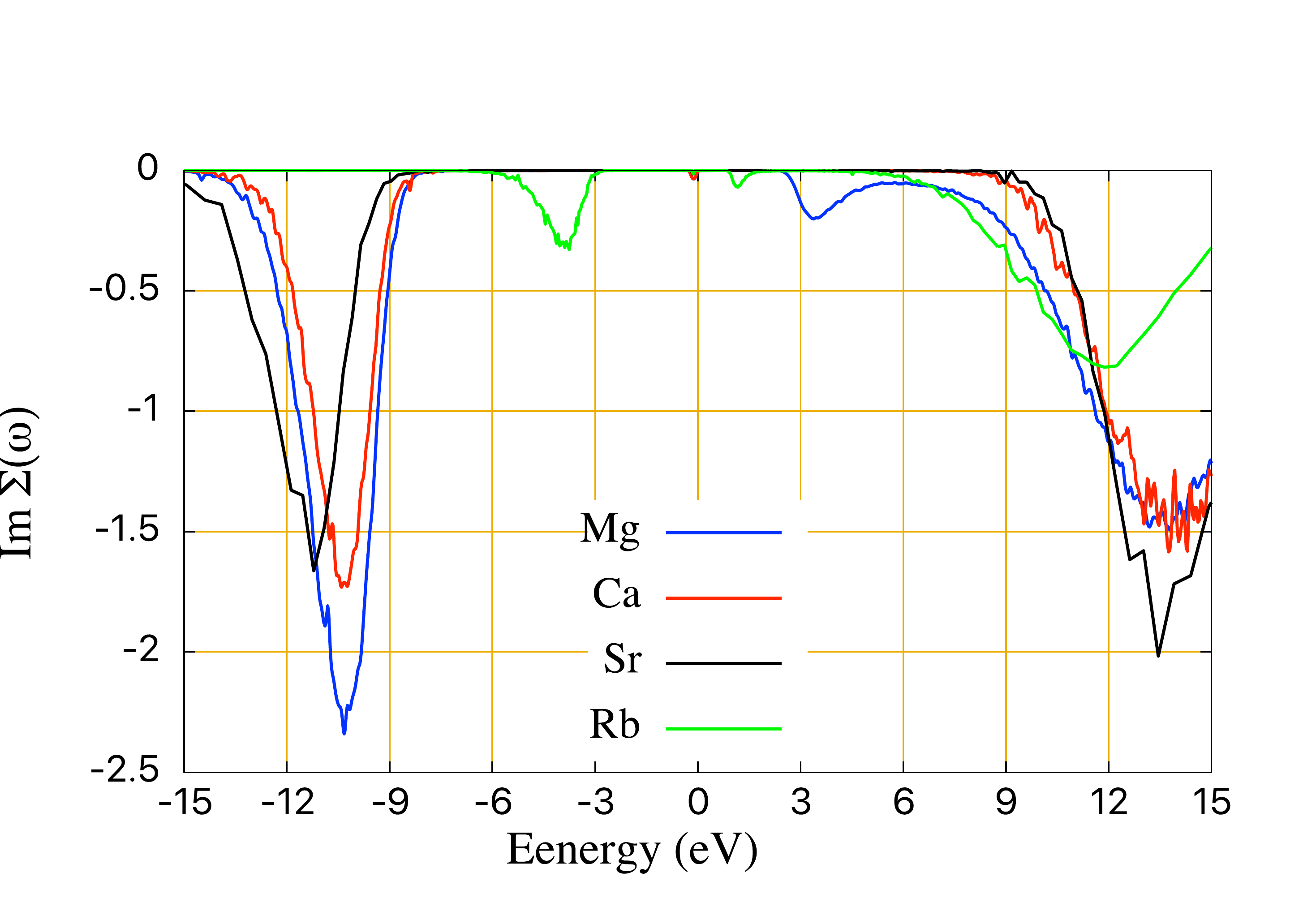}
\caption{ (Color online)
Imaginary part of the eDMFT self-energy of Mg, Ca, Sr, and Rb
}
\end{figure*}

In Supplementary Figures 3-7, we describe full band structures for elemental Na, K, Mg Be, Sr, Cs, Ca, and Rb respectively as computed in LDA, mBJ, G$_0$W$_0$, B3LYP, and eDMFT in various energy windows. They clearly show the differences in bandwidths for occupied as well as unoccupied bands. The renormalization of the bands also shows a strong K-dependence as the differences in bandwidth among various methods vary strongly in various parts of the Brillouin zone. The difference is always largest at the high-symmetric $\Gamma$-point. We find B3LYP not only overestimates bandwidths of occupied bands, but it also does the same for the unoccupied bands. Again, LDA and mBJ bands disperse very similarly. For K, Sr, Cs we notice the bandwidths are reduced in GW compared to LDA for occupied bands but unoccupied bands, the bandwidths in GW are larger than that of LDA or mBJ.

\vspace{5in}



\begin{figure*}
\includegraphics[width=440pt, angle=0]{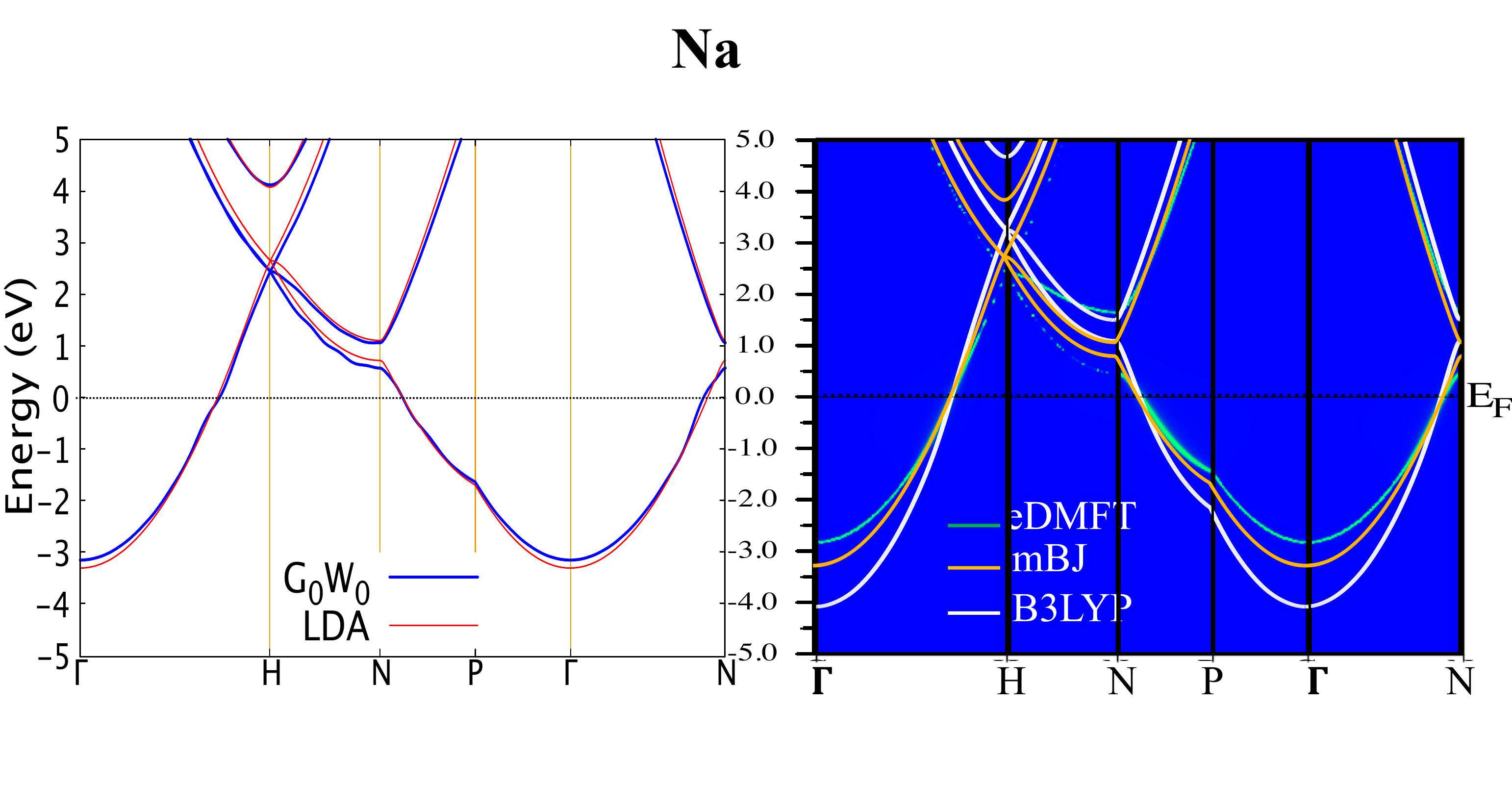}
\includegraphics[width=440pt, angle=0]{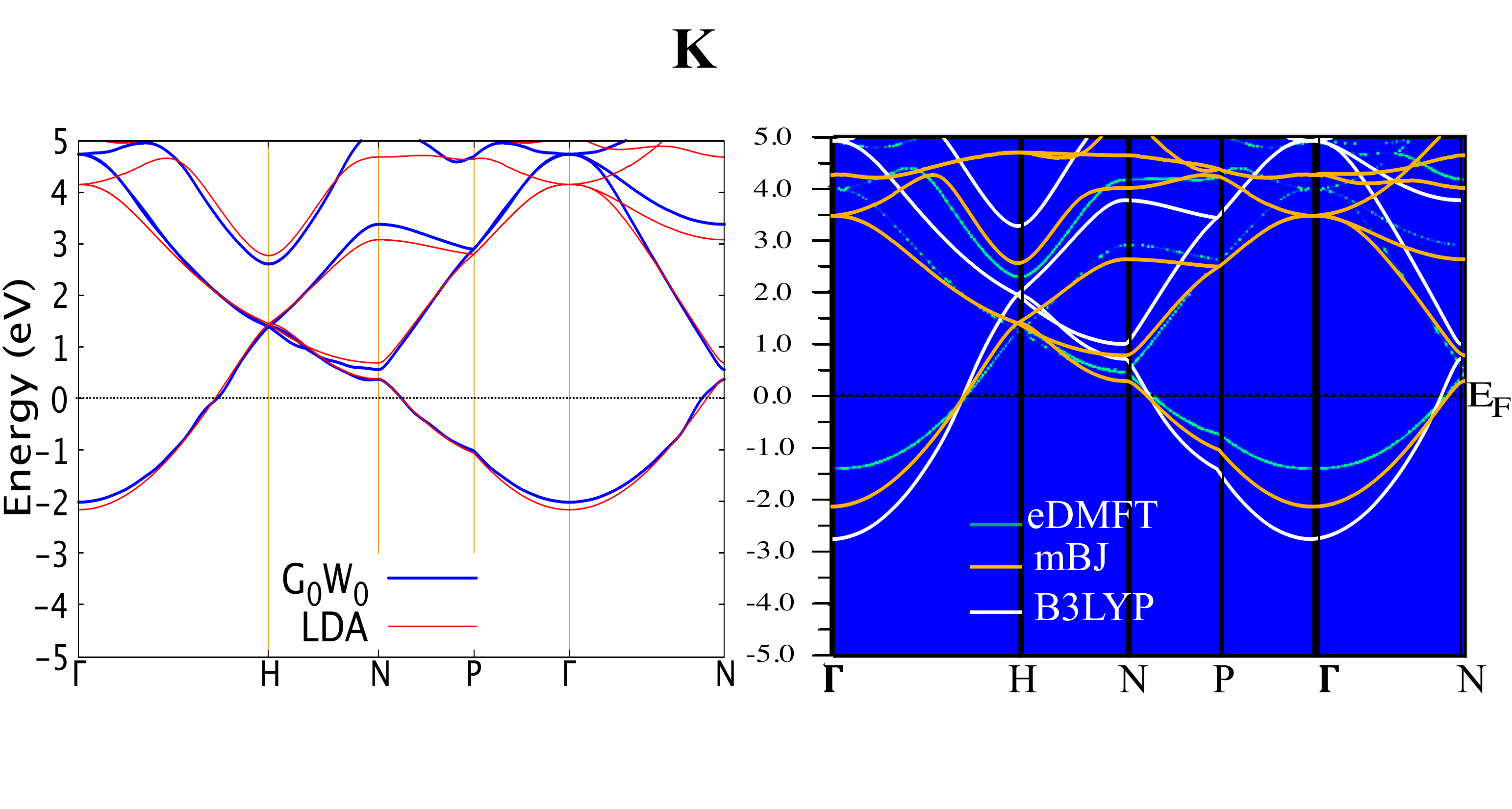}
\includegraphics[width=440pt, angle=0]{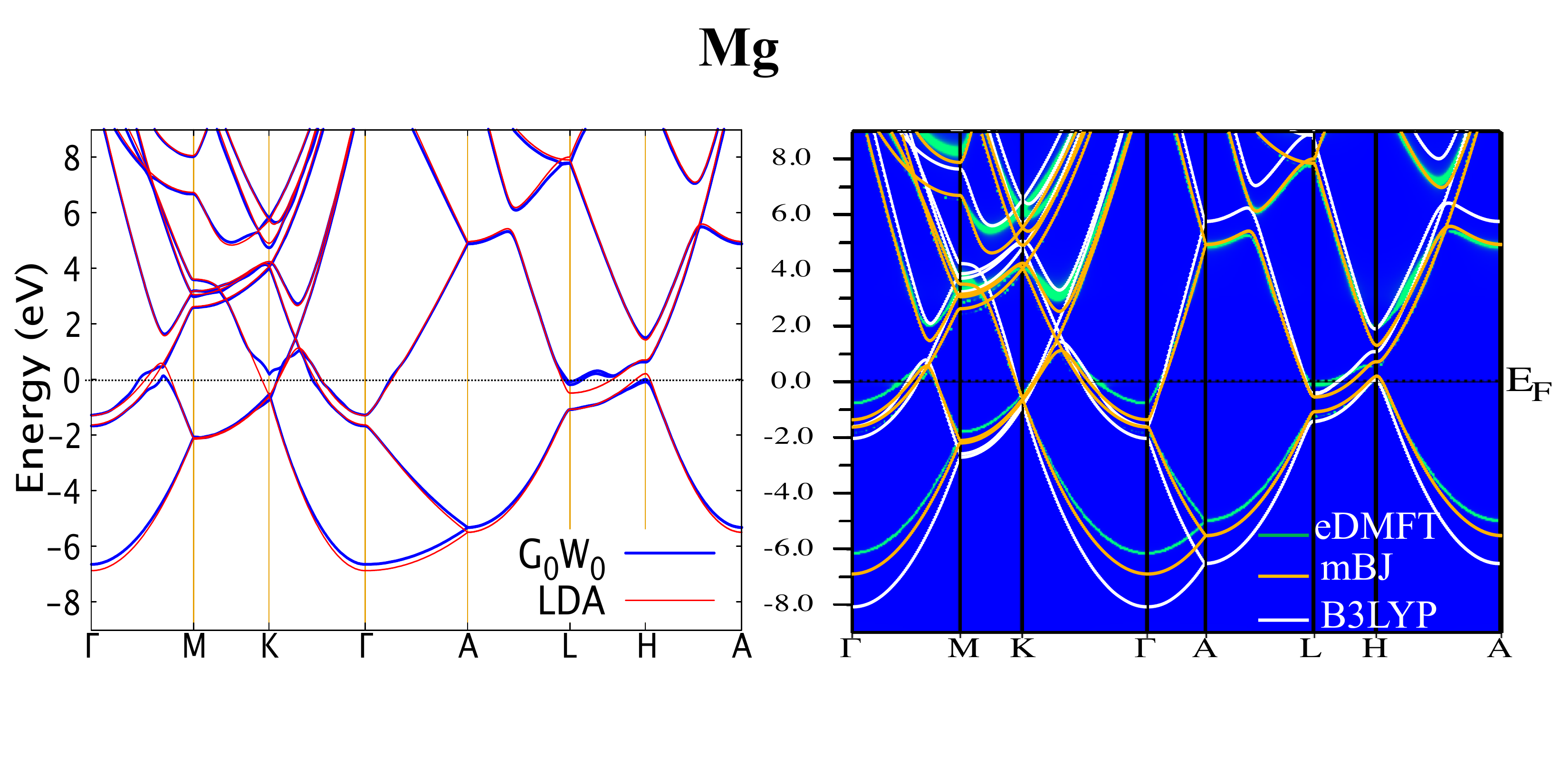}
\caption{Full band structures of elemental Na (top), K (middle) and Mg (bottom) as computed in LDA, mBJ, G$_0$W$_0$, B3LYP, and eDMFT.
}
\end{figure*}

\begin{figure*}
\includegraphics[width=440pt, angle=0]{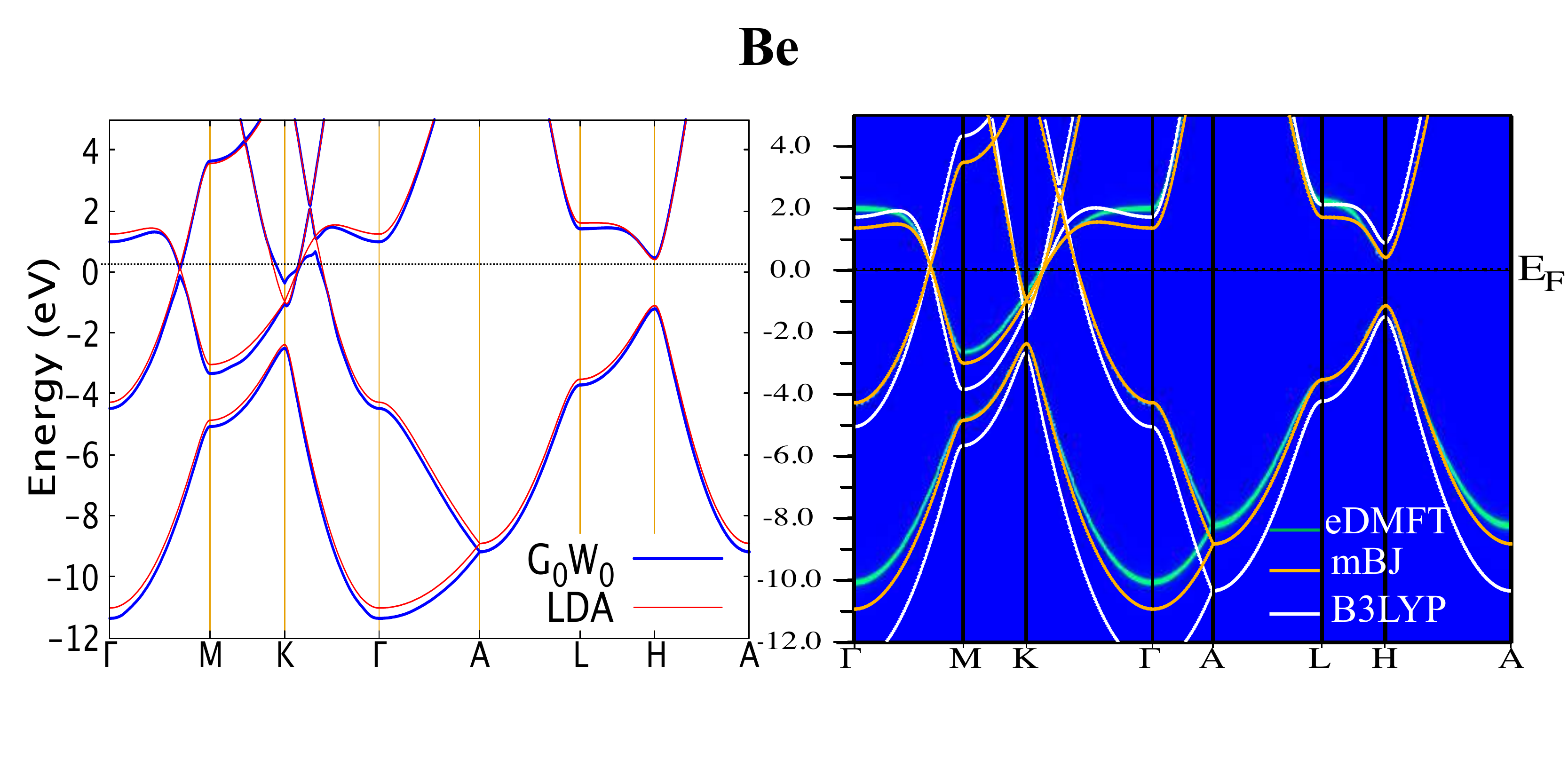}
\includegraphics[width=440pt, angle=0]{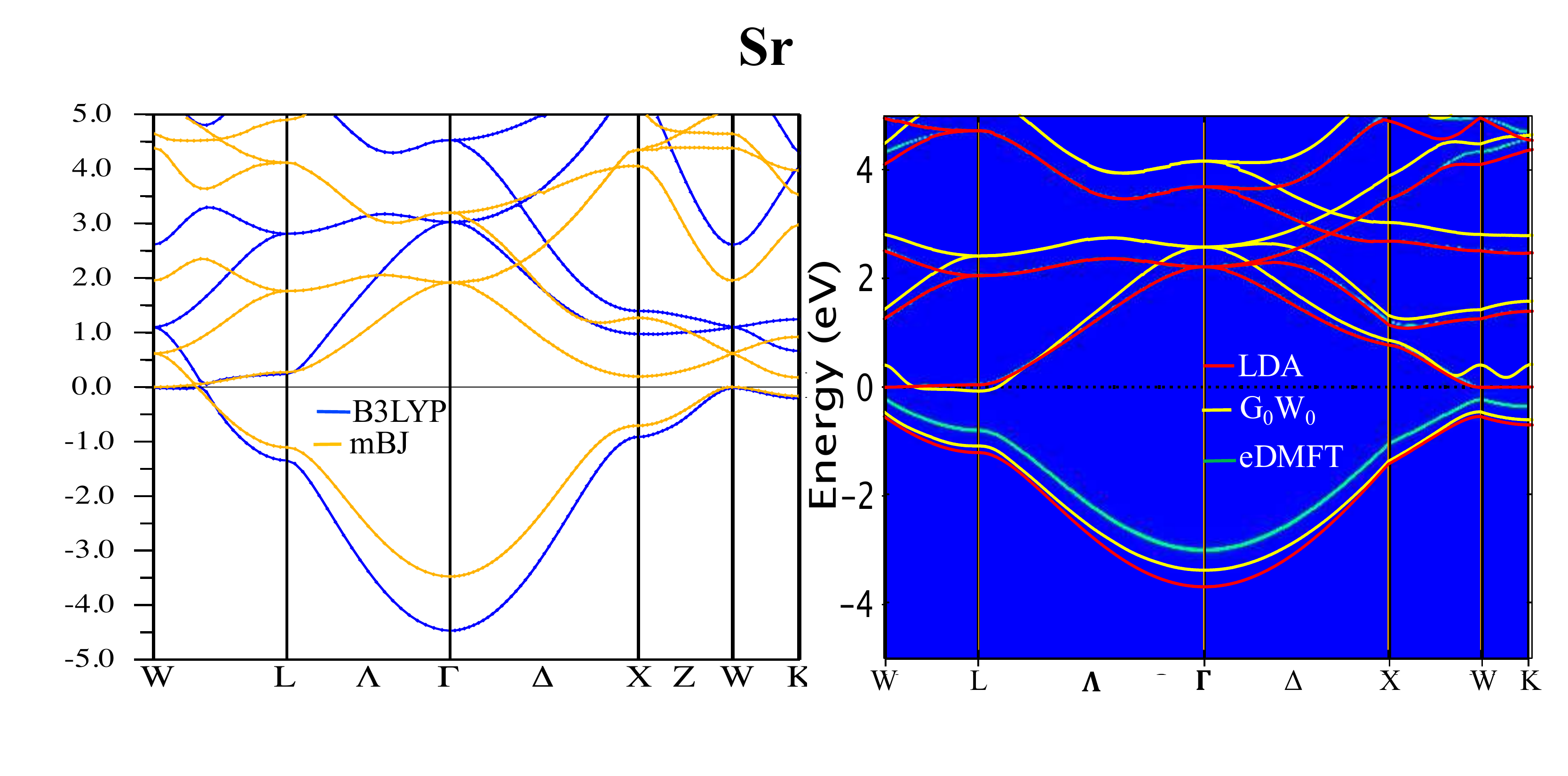}
\includegraphics[width=440pt, angle=0]{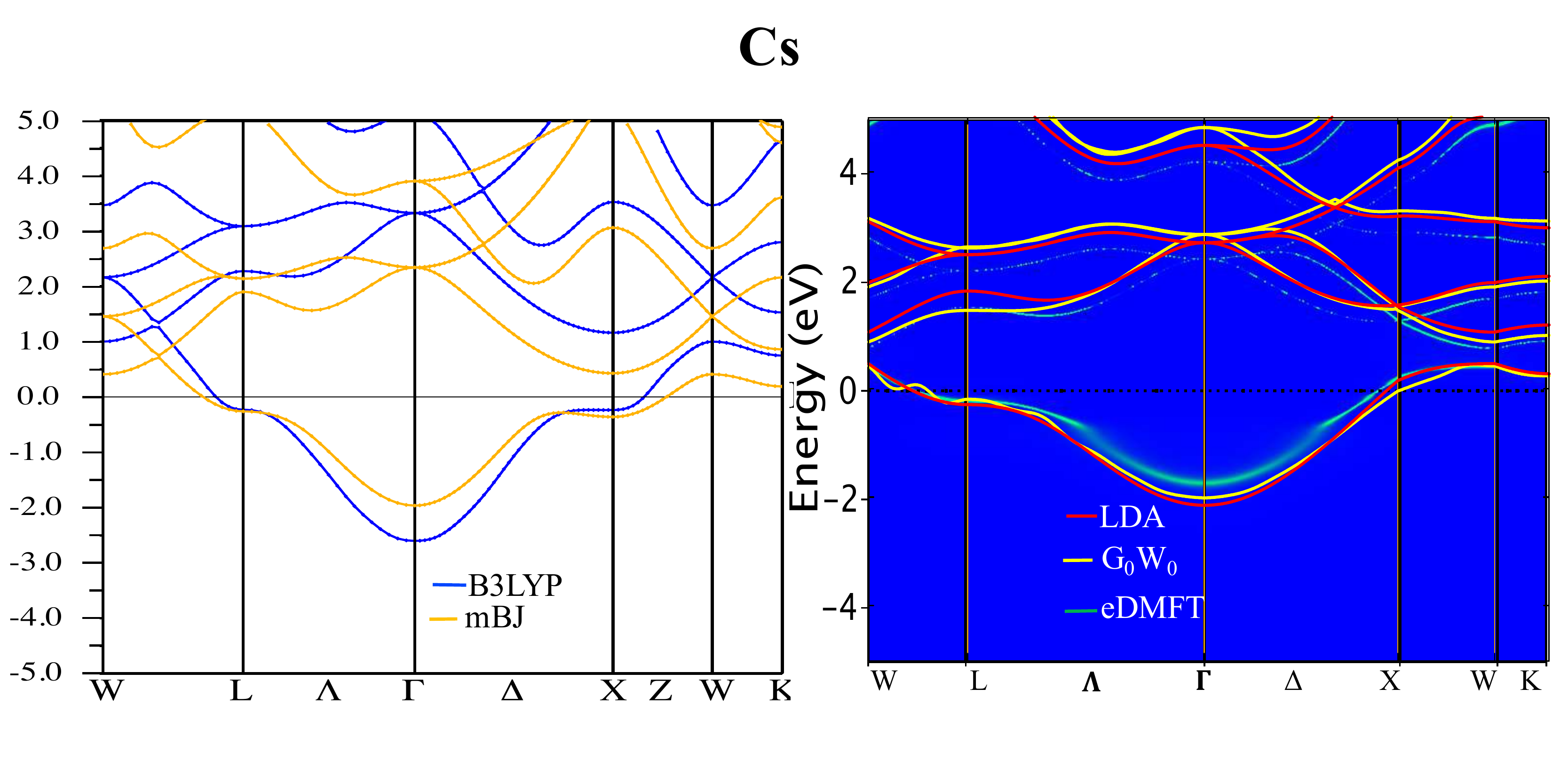}
\caption{ (Color online)
Band structures of elemental Be (top), Sr (middle) and Cs (bottom) as computed in LDA, mBJ, G$_0$W$_0$, B3LYP, and eDMFT.
}
\end{figure*}

\begin{figure*}
\includegraphics[width=480pt, angle=0]{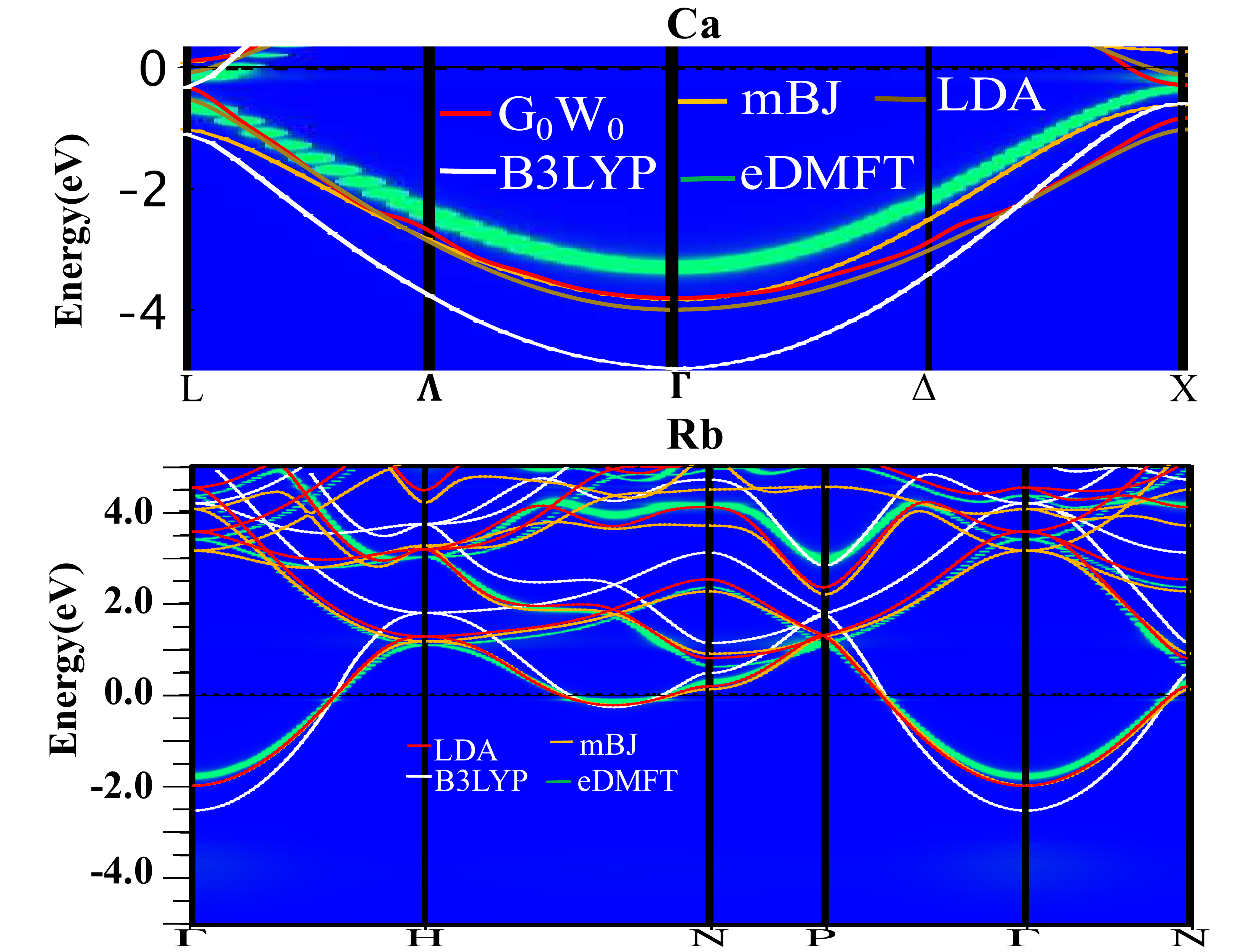}
\caption{ Band structures of elemental Ca (top) and Rb (bottom) as computed in LDA, mBJ, G$_0$W$_0$, B3LYP, and eDMFT.
}
\end{figure*}

\newpage



\end{document}